\documentclass[aps,pre,eqsecnum,showpacs,draft]{revtex4} 
\usepackage{amsmath,bm}

\begin{document}

\draft 
\title{Quasistationary states and coherent versus incoherent transitions for
crossing diabatic potentials.}
 
\author{V.A.Benderskii} 
\affiliation {Institute of Problems of Chemical Physics, RAS \\ 142432 Moscow
Region, Chernogolovka, Russia} 
\affiliation{Laue-Langevin Institute, F-38042,
Grenoble, France} 
 
\author{E.V.Vetoshkin} 
\affiliation {Institute of Problems of Chemical Physics, RAS \\ 142432 Moscow
Region, Chernogolovka, Russia} 
\author{E. I. Kats} \affiliation{Laue-Langevin Institute, F-38042,
Grenoble, France} 
\affiliation{L. D. Landau Institute for Theoretical Physics, RAS, Moscow, Russia}
 
\date{\today}

\begin{abstract}
We investigate  coherent and incoherent tunneling phenomena
in conditions of crossing diabatic potentials. 
We consider 
a model of two crossing parabolic diabatic potentials (left (L) and right (R))
with an independent of
coordinates 
constant adiabatic coupling $U_{12}$. 
As a result of the coupling and level crossing avoiding, we get the asymmetric
double-well lower adiabatic potential with a variable shape 
depending on a value of a continuous parameter $b$ (which describes in the limit
$b=1$ two identical parabolic diabatic potential crossing and in the limit
$b \to \infty $ one-well and linear diabatic
potentials crossing).                                                                                         
We show that the doublet structure
of levels (generic for double-well potentials) is remained valid as
long as the transition matrix element $H_{LR}$ (or tunneling splitting)
is smaller than characteristic inter-level spacings $\Delta _R$
(the latter ones in own turn decrease upon increasing
of a difference between the diabatic potential minima $\delta E_{LR}$).
We calculate the non-adiabatic factor, i.e. $H_{LR}$ as a function of $U_{12}$.
In the diabatic limit 
($U_{12} \to 0$) 
$H_{LR}$ goes to zero, and
in the adiabatic limit ($U_{12} \to \infty $) 
the tunneling
transitions do not depend on the upper potential.
In the over-barrier energy region $H_{LR}$ is an oscillating function
of $U_{12}$, due to the resonances between the states in the lower and in the upper
adiabatic potentials. 
In the case $H_{LR} > \Delta _R$, any level from the shallow $L$-well is coupled
by the tunneling to several levels in the $R$-well,
and the transitions lose their coherence.
A new phenomenom emanated from this oscillating dependence of $H_{LR}$ on $U_{12}$,
namely, multiple coherent - incoherent regime 
transitions
for the upper adiabatic potential state evolution, is our main concern in this paper.
The problem
is not only of intellectual interest but also of
relevance to various molecular
systems undergoing conversion of electronic states or
isomerization reactions. Our model exhausts all cases practically relevant for
spectroscopy of non-rigid molecules, and 
can capture many of the features
exhibited by experiment.                                                                                       
\end{abstract}

\maketitle
\section{Introduction}
\label{I}

Double-well potentials appear in various contexts in physics and chemistry.
For example the simplest pattern of almost any molecular reactive
system (with two stable configurations identified as a reactant and as a product)
corresponds to the model potential energy formed by two multidimensional
more or less parabolic surfaces shifted relative to each other.
Although the 1D asymmetric double-well model is idealized, it can be very useful
for a qualitative discussion to gain more insight into complex
multidimensional dynamic molecular properties for which exact or even approximate theoretical
results are not available, thus 
throughout what follows we will consider $1D$ case only.  

In the classical limit for the energy region $E < U_b$ (where $U_b$ is the potential barrier,
separating the both, say $L$ and $R$ wells) which will be referred further on as the tunneling region,
the behavior in the both wells are fully decoupled and therefore 
one well is independent from
other. 
As it is a common wisdom nowadays in quantum mechanics even for $E < U_b$ 
the particle can tunnel
between the wells. 
It admixes the L and R well localized states, thus allowing an under-barrier
tunneling mechanism. The extent of this delocalization is larger in the states close to
the top of the barrier, and it is maximal when the unperturbed levels
on the opposite sides of the barrier are degenerate (the reason is
immediately clear by looking at the standard textbook expressions
for the tunneling probability and splitting \cite{LL65}).
For the symmetric case this tunneling 
level splitting leads to coherent quantum oscillations
typical for any two-level system. 
For asymmetric double well potentials, 
pairs are not in coincidence any more, the tunneling is suppressed,
except for certain critical values
of model parameters for which the levels are brought in resonance again,
and the problem becomes more tricky.
We have recently shown \cite{BK02} that one can successfully attack
this problem by a semiclassical solution of the Schr\"odinger equation for
1D asymmetric double well potential with one-parameter dependent shape
\begin{eqnarray}
\label{a2}
U_1(X) = \frac{1}{2}X^2(1-X)\left (1 + \frac{1}{b^2}X\right )
\, .
\end{eqnarray}
$U_1$ is written in dimensionless form measuring energy in the 
characteristic frequency of the
oscillations say around the left (L) minimum $\Omega _0$, and the coordinate $X$
is measured in the units $a_0$ of the inter-well distance (and in ''God given unit''
we put $\hbar = 1$, except where explicitely stated to the contrary
when it is necessary for understanding or estimations). The dimensionless parameter $b$
allows us to change the shape of the
$R$ well and to consider both limiting cases, namely a traditional
symmetric double-well potential ($b=1$), and a decay potential for $b \to \infty $.
Behavior in the latter limiting case is also well known, there is a continuum spectrum of
eigenstates for $X \to \infty $ and incoherent decay of quasi-stationary
states from the $L$ - well.

For a small asymmetry $b \simeq 1$ the ground state doublet $E_0^\pm $ in the potential (\ref{a2})
\begin{eqnarray}
\label{a4}
E_0^\pm = \pm \left [\frac{(b-1)^2}{4} + \left (H_{LR}^{(0)}\right )^2\right ]^{1/2}
\, ,
\end{eqnarray}
where $H_{LR}^{(0)}$ is the ground state splitting for the
symmetrical double well potential.

It was shown in \cite{BK02} that 
for any asymmetric double-well potential
the behavior depends crucially
on a dimensionless parameter $\Lambda $ that is, roughly speaking,
a ratio of characteristic frequencies for low-energy in-well oscillations
and inter-well tunneling. 
For $\Lambda \ll 1$, there are well defined resonance
pairs of levels, and so-called survival probability (i.e. the probability
for a particle initially localized in one well to remain there) has
coherent oscillations related to resonance splitting.
However, for $\Lambda \to \infty $ for any finite
time scale, there are no oscillations for the survival probability, and there
is almost an exponential decay with the characteristic 
relaxation time $\propto H_{LR}^{-2}$  
determined by Fermi golden rule. 
In the NMR language this relaxation time can be associated to
the so-called dephasing $T_2$ time.
Thus one can say that tunneling destroys coherent
behavior and it can be associated with dephasing
processes in the phenomenological Bloch theory
of quantum relaxation. Explicitely for asymmetric double-well
potentials
\begin{eqnarray}
\label{I1}
T_2^{-1} = \frac{H_{LR}^2}{2\pi \Delta _R}
\, ,
\end{eqnarray}
where $\Delta _R$ is a typical level spacing for the final states.
In the case $\Lambda \gg 1$ one may not restrict himself to
the only resonance pair levels. The number of levels perturbed
by tunneling grows proportionally to $\sqrt \Lambda $, or, 
by other words, instead of isolated pairs there appear the
resonance regions containing the sets of strongly coupled levels.
At the intermediate values of $\Lambda \simeq 1 $ one has a crossover between
both limiting cases, namely the exponential decay with subsequent
long period recurrent behavior (more longer the larger $\Lambda $).
In this region, $\Lambda \simeq 1$, and in the vicinity
of quasi-stationary states of the $L$- well the doublet-like eigen-spectrum turns into the equidistant one.

It is particularly instructive to look to this result from a slightly different
point of view related to a striking and still enigmatic 
phenomenom - quantum chaos. Perhaps the first successful quantitative
criteria relating the classical ergodic theory to
quantum molecular dynamics was formulated long ago by von Neumann and Wigner
\cite{NW29}. According to \cite{NW29} a system has the ergodic behavior
if it has:

(i) equidistant spectral distribution (i.e. no degenerate states);

(ii) time decay of correlations for any observable.

Let us emphasize that the criteria claim that quantum manifestations
of the classical chaos are related to specific spectral features of a system
under consideration, and not to some kind of its
irregular time evolution. From the first sight it might seem that ergodic behavior
is impossible in one dimensional systems which are integrable in classical mechanics.
However, in fact our model examples should be considered as two classically decoupled
systems ($L$ and $R$ wells) interacting only via quantum tunneling. One can call this phenomenom
as tunneling induced ergodicity, and for more details, see e.g. \cite{BK02}, (and also,
\cite{BT93}, \cite{KK03}).
It turns out that the time evolution of the localized initial state is governed
by the interplay of two physical parameters, recovering period time and decay rate.
The both parameters depend on the spectral representation (or distribution function) of the
initial state.
It is shown in \cite{BK02} that the both listed above criteria are satisfied for strongly
asymmetric double-well potentials where highly excited states
in the $R$ well are strongly perturbed by tunneling
from the shallow $L$ well.
The condition to have this phenomenon (tunneling induced ergodicity of final
states) reads in our notation as
 
\begin{eqnarray}
\label{na1}
|H_{LR}| > \Delta _R
\, .
\end{eqnarray}
In this paper we will show that similar phenomena can take place also
for delocalized initial states, where non-adiabatic coupling gives rise
chaotic behavior. It will be referred in what follows as 
non-adiabatic transitions induced ergodicity.

Of course an isolated double-well potential is only an idealization
of any real molecular system. The applicability of such an idealization
must be decided separately for each system or process in question.
However even in the cases where such a model is not justifiable
the calculations we performed are nonetheless instructive.
But in this paper we make one step further.
In a typical problem of chemical dynamics or molecular spectroscopy,
the double-well potentials can appear as a result of level crossing phenomena,
and the consideration of only the isolated double-well potential
(lower adiabatic potential) can be justified only if
the gap occurring in the spectrum at the avoided level crossing
point is much larger than all other characteristic energy scales of
the problem.
However, evidently it is not the case
for example if we are interested in the calculation of vibrational - tunneling spectra of non-rigid
molecules, or reactive complexes with more than one stable configuration. The lowest 
multi-well potential
of such systems is formed from one well diabatic potentials crossing corresponding to each stable
configuration. Apart from the lowest potential, the upper adiabatic potential with its minimum above the
maximum of the lowest potential should be also taken into account for these situations (see Fig. 1). In
the most of the calculations of tunneling splittings in the ground and low excited vibrational states the
coupling to the upper potential are neglected, what is certainly correct only for strong enough adiabatic
coupling. The same situation takes place for systems undergoing the Jahn - Teller effect,
where the interference of the diabatic states occurs \cite{BE84}.
In all these situations the adiabatic coupling removes diabatic level crossing, and the diabatic
levels are replaced by the adiabatic ones. Let us repeat that
only in the case of a large adiabatic splitting 
one can restrict oneself to the only lower adiabatic potential
and neglect
any influence of the upper adiabatic potential.
However, in a general case of arbitrary adiabatic splittings,
intra-well and inter-wells dynamics depends on the both adiabatic potentials
(i.e. on tunneling splittings
and adiabatic interactions).

In the fundamental problems of chemical dynamics and molecular spectroscopy, 
the transitions from the initial to final states can be treated as 
a certain motion along the
potential energy surfaces of the system under consideration.
These surfaces in own turn are usually determined within
the Born - Oppenheimer approximation.
However, the approximation becomes inadequate for the excited vibrational
states, when their energies are of the order of electronic inter level energy
spacing or near the dissociation limit. In the both cases the non-adiabatic
transitions should be taken into account, and the most of the non-radiative
processes occur owing to this non-adiabaticity. The typical examples
investigated in the monography \cite{EL83}, are so-called pre-dissociation,
singlet-triplet or singlet-singlet conversion, and vibrational relaxation
phenomena. 

To treat this kind of level crossing (Landau - Zener (LZ))
problems usual textbook consideration utilizes the outset within
a limited electronic subspace which is completely
spanned by a finite set of Born - Oppenheimer or adiabatic
electronic states. However, because these states obey
the noncrossing rule it may be desirable
technically to transform the states into the diabatic
representation in which the diagonal matrix elements
of the electronic Hamiltonian in the subspace
can cross, and the off-diagonal interactions
appear as scalar coupling potentials.

The major concern of this paper is with the construction and solution
of a model
for two asymmetric diabatic level crossing phenomena. 
The rest of our paper begins in section \ref{bas} with a 
formulation of our model and with a
discussion of basic methodical details
necessary for our study.
Section \ref{res} contains our main
results. We derive the criteria
for reversibility and coherent or incoherent tunneling for  
crossing diabatic potentials. 
Our conclusion section \ref{con} deals with miscellaneous subjects related
to the diabatic level crossing phenomena.

\section{Model potential and basic relations}
\label{bas}     

As a model for diabatic potentials in this paper we choose two non-equivalent
parabola
\begin{eqnarray} 
\label{nn1} 
U_L = \frac{1}{2}\left (1 + X\right )^2\, ; \,
U_R = \frac{1}{2}\left (1 - 2 X + \frac{X^2}{b}\right )
\, 
\end{eqnarray} 
with a symmetrical crossing in the point $X=0$.
Upon increasing the well asymmetry
\begin{eqnarray} 
\label{nn2} 
\delta E_{LR} = - \frac{b-1}{2}
\,  
\end{eqnarray} 
the potential $U_R$ is converted from a simple
parabola at $b=1$ to a linear potential at $b \to \infty $.
Owing to the adiabatic coupling $U_{12}$ (which
we assume for simplicity independent of coordinates) we get the lower
double-well and the upper one-well adiabatic potentials (see Fig. 1). 

At arbitrary values of the parameters $U_{12}$ and $b$ to find
eigenstates and eigenfunctions for our model potential we
should solve the coupled Schr\"{o}dinger equations
\begin{eqnarray} 
\label{nn3} 
-\frac{d^2 \Theta _L}{d X^2} + \gamma ^2 (U_L(X) - E) \Theta _L = \gamma ^2 U_{12} \Theta _R
\, ; \,
-\frac{d^2 \Theta _R}{d X^2} + \gamma ^2 (U_R(X) - E) \Theta _R = \gamma ^2 U_{12} \Theta _L
\, ,
\end{eqnarray} 
which can be written as one fourth order equation
\begin{eqnarray} 
\label{nn4} 
\frac{d^4 \Theta _L}{d X^4} - \gamma ^2 (U_L(X) + U_R(X) -2E)\frac{d^2 \Theta _L}{d X^2} 
- 2 \gamma ^2 \frac{d U_L}{d X}\frac{d \Theta _L}{d X} + 
\gamma ^4 \left [(U_L - E)(U_R - E)
- U_{12}^2 - \frac{1}{\gamma ^2}\frac{d^2 U_L}{d X^2}\right ] \Theta _L = 0
\, .
\end{eqnarray} 
Here
$\gamma \gg 1$ is the semiclassical parameter which is determined by the ratio
of the characteristic potential scale over the zero oscillation energy (i.e. as
above $\gamma \equiv m \Omega _0
a_0^2/\hbar $, where
$m$ is a mass of a particle, $a_0$ is a characteristic length of the problem, e.g. the tunneling
distance, $\Omega _0$ is a characteristic frequency, 
e.g. the oscillation frequency around the potential minimum).
 
Luckily the equation (\ref{nn4}) admits semiclassical solutions
by Fedoryuk method \cite{FE64} - \cite{FE66} since the coefficients at the
$n$-th order derivatives proportional to $\gamma ^{-n}$, and therefore are small.
Besides in the vicinity of the crossing point $X=0$ the diabatic potentials
(\ref{nn1}) can be replaced by the linear ones counted from the barrier top
$U^\# $
\begin{eqnarray} 
\label{nn6} 
U_{L/R}(X)  = U^\#  \pm f X
\, ,
\end{eqnarray} 
and eventually the equation (\ref{nn4}) can be presented into a more compact
and simple form
\begin{eqnarray} 
\label{nn7} 
\frac{d^4 \Theta _L}{d X^4} - 2\gamma ^2 \alpha \frac{d^2 \Theta _L}{d X^2}
- 2 \gamma ^2 f \frac{d \Theta _L}{d X} 
+ \gamma ^4 [\alpha ^2 - f ^2X^2 - U_{12}^2] \Theta _L = 0
\, ,
\end{eqnarray} 
where $\alpha = U^\# - E$.

Four roots of the characteristic polynomial of (\ref{nn4}) or (\ref{nn7})
\begin{eqnarray} 
\label{nn8} 
F(\lambda , X) = \lambda ^4 - \gamma ^2 (U_L + U_R - 2 E)\lambda ^2
- 2\gamma ^2 \frac{dU_L}{d X}\lambda + \gamma ^4 
\left [(U_L - E)(U_R - E)
- U_{12}^2 - \frac{1}{\gamma ^2}\frac{d^2 U_L}{d X^2}\right ]
\, 
\end{eqnarray} 
determine the four fundamental solutions to (\ref{nn7})
\begin{eqnarray} 
\label{nn9} 
y_j = (f^2X^2 + U_{12}^2)^{-1/4}\exp \left (\int \lambda _j (X) d X\right ) \, , \, j = 1 , 2 , 3 , 4
\, .
\end{eqnarray} 
The solutions (\ref{nn9}) can be visualized as a motion with imaginary
momenta in the upper and lower adiabatic potentials
\begin{eqnarray} 
\label{nn10} 
U^\pm = \frac{1}{2}(U_L + U_R) \pm \frac{1}{2} [(U_L - U_R)^2 + 4 U_{12}^2]^{1/2}
\, . 
\end{eqnarray} 

As it was mentioned above in the vicinity of the crossing point one can replace (\ref{nn4})
by (\ref{nn7}). In the latter equation the coefficient at the first order derivative
is small ($\propto \gamma ^{-1}$), and by the substitution
\begin{eqnarray} 
\label{nn11} 
\Theta _L = \exp (\kappa _{1 , 2} X) \Phi _L^{1 , 2}
\, ,
\end{eqnarray} 
where
\begin{eqnarray} 
\label{nn12} 
\kappa _{1 , 2} = \pm \gamma \sqrt \alpha \left (1 \pm \frac{\delta }{2} \right )
\, , 
\end{eqnarray} 
and $\delta $ is a first order correction (see \cite{BV03})
$ \delta = (\gamma f/4 \sqrt {\alpha })$.
Therefore
the equation (\ref{nn7}) is reduced to two independent Weber equations
with the known fundamental solutions \cite{EM53} 
\begin{eqnarray} 
\label{nn13} 
\{\Theta _L\} = 
\left \{ 
\exp( \pm \gamma \sqrt \alpha X) D_{-\nu } \left (\pm \left 
(\frac{\gamma ^2f^2}{\alpha }\right )^{1/4}X\right ) \, , \, 
\exp( \pm \gamma \sqrt \alpha X) D_{-1 - \nu } \left (\pm \left 
(\frac{\gamma ^2f^2}{\alpha }\right )^{1/4}X\right ) \right \}
\, , 
\end{eqnarray} 
where $\nu \equiv (\gamma U_{12}^2/4 f\sqrt \alpha )$ is referred traditionally as the 
Massey parameter, and in fact it controls the main features of the behavior.
The corrections to the indices of the parabolic cylinder
functions $D$ and to the arguments of these functions can be 
found from (\ref{nn12}) and have been calculated in \cite{BV03}.

At the next step 
we should perform the asymptotically smooth matching of the solutions
(\ref{nn9}) and (\ref{nn13}).
The whole analysis can be brought into a more elegant form
by introducing connection matrices which link
on the complex plane the semiclassical solutions to the
Schr\"odinger equation for the exact potential of the problem under
study (e.g. (\ref{nn1}) for our case) and the exact solutions
of the so-called comparison equation (in our case (\ref{nn7}))
which is valid near the crossing point.
The explicit calculations of the connection matrices are rather
involved since the LZ problem is characterized by the four fundamental
solutions to the left and to the right regions with respect to turning or crossing
points. Therefore the connection matrices, we are looking for,
are $4 \times 4$ matrices. Although the generalization for our case of
the known already $2 \times 2$ connection matrices (see e.g.,
\cite{HE62}, and for more recent references our publication \cite{BV03})
is straightforward, it deserves some precaution as it implies quite
different procedures for the energy, more accurately for $E/\gamma $
smaller (the tunneling region), larger (the over-barrier region),
or of the order (the intermediate region) of the potential barrier, i.e.
$U^\# - U_{12}$.

Indeed, in the case 
\begin{eqnarray}
\label{j1}
\frac{E}{\gamma }\ll U^\# - U_{12} 
\, ,
\end{eqnarray} 
the region near the crossing point
is forbidden for the both adiabatic potentials. However, four real-valued
turning points of the lower adiabatic potential are far enough from the crossing point.
The upper adiabatic potential in this case is also  
higher than $E/\gamma $, and therefore for the instanton approach there are
two imaginary turning points which characterize the motion in the inverted
upper adiabatic potential. Thus for the tunneling region
we have four real-valued and two pure imaginary turning points.

In the over-barrier energy region, when the energy is larger
than the upper adiabatic potential minimum, i.e.
\begin{eqnarray}
\label{j2}
\frac{E}{\gamma } \gg U^\# + U_{12} 
\, ,
\end{eqnarray} 
the whole region for the both potentials, is accessible
for the
classical motion. Therefore there are four real-valued turning points (two for
the lower and two for the upper adiabatic potentials). Besides there are two
imaginary turning points corresponding to the quantum over-barrier reflection for the
lower adiabatic potential.
Finally in the intermediate energy region, i.e. for 
\begin{eqnarray}
\label{j3}
U^\# + U_{12} \geq \frac{E}{\gamma }\geq
U^\# - U_{12}  
\, ,
\end{eqnarray}
there are two real-valued and four imaginary turning points.

The tunneling path is one central point to be considered within the
instanton method, and the determination of the tunneling
trajectory (or trajectories) is, in a general case,
a nontrivial task. However for our model 1D potential
(\ref{a2}) in the symmetrical case 
the extremal action trajectory consists from so-called kink and anti-kink
pairs corresponding $L \to R$ and $R \to L$ transitions, and the action for 
every part (i.e. kink or anti-kink) is $W^*$. More or less qualitatively the same
is the tunneling path for a small potential asymmetry. However, when the asymmetry
is larger than the tunneling splitting in the symmetric double-well
potential, there is only one classical trajectory starting from the less
deep  well (say $L$) which does not reach the more deep $R$ minimum and
comes back to $L$. Thus in this case the pair kink - anti-kink forms
a single so-called bounce trajectory with the action $2W^*$. We will explore
this issue in more details in what follows.

The double-well shape of the lower adiabatic potential and the influence 
of the upper adiabatic
potential require that to find the solutions one has to take into account
at least two instanton trajectories with the energies $E=0$ and $E=\gamma V^\# $.
Following 
the strategy, described above, one has to match smoothly the semiclassical
(e.g., instanton) solutions known in the remote 
from the crossing point ($X=0$) region with the solutions of the more simple
comparison equation which is valid in the vicinity of the crossing point.
This matching should be performed asymptotically, i.e. at small $|X|$ but for
large enough $\sqrt \gamma |X|$.

Now we are in the position to find all needed connection matrices.
In the tunneling region (\ref{j1}) for every well ($L$ or $R$) there
exist increasing and decreasing exponentially real-valued solutions to the Schr\"odinger
equation. The solutions are matched at the crossing point, therefore
they are linked by the real-valued $4 \times 4$ connection matrix which
should have two $2 \times 2$ blocks linking the increasing (decreasing)
diabatic solution in the $L$-well with the decreasing (increasing)
diabatic solution in the $R$-well, in the agreement with the standard Landau
scheme of the tunneling transitions \cite{LL65}.
Omitting a large amount of tedious algebra we can represent the connection matrix
linking the ''asymptotic'' (i.e. in the left/right ($L$, $R$) wells and for
the upper/lower ($+$, $-$) adiabatic potentials) solutions 
in the tunneling energy region in the following form 
\begin{eqnarray} && 
\label{q1}
\left ( 
\begin{array}{c} 
\Phi _R^- \\ 
\Phi _R^+ \\ 
\Phi _L^+ \\ 
\Phi _L^- 
\end{array} 
\right ) = 
\left ( 
\begin{array}{cc} 
\hat {M}_c^{(+)}\hat {L}^{(c)}_R\hat {M}_c^{(-)}\hat F_c & 0 \\ 
0 & \hat 1 
\end{array}
\right )
\hat {U}_c 
\left ( 
\begin{array}{cc} 
\hat F_c\hat {M}_c^{(+)}\hat {L}^{(c)}_L\hat {M}_c^{(-)} & 0 \\ 
0 & \hat 1 
\end{array} 
\right )
\left ( 
\begin{array}{c} 
\Phi _L^+ \\ 
\Phi _L^- \\ 
\Phi _R^- \\ 
\Phi _R^+ 
\end{array} 
\right ) \, .
\end{eqnarray} 
Here $\hat U_c$ is the $4 \times 4$ connection matrix at the crossing
point, which in the tunneling region has the following form 
\begin{eqnarray} && 
\label{q8} \hat {U}_c =
\left [
\begin{array}{cccc} 
p & 0 & 0 & -\cos (\pi \nu ) \\ 
0 & (\sin^2 (\pi \nu ))/p & - \cos (\pi \nu ) & 0 \\ 
0 & \cos (\pi \nu ) & p & 0 \\ 
\cos (\pi \nu ) & 0 & 0 & (\sin ^2(\pi \nu ))/p 
\end{array} 
\right ] \, , 
\end{eqnarray} 
where we designated 
\begin{eqnarray}
\label{xx}
p=\frac{\sqrt {2\pi }\exp(-2 \chi )}{\Gamma (\nu )}
\, ,
\end{eqnarray}
and
$ \chi =
(\nu /2) - (1/2)\left (\nu - (1/2) \right ) \ln \nu $. 
The matrices $\hat {M}_c^{(+)}$ and $\hat {M}_c^{(-)}$
are the $ 2 \times 2$ connection matrices at the corresponding linear turning points,
which are determined by the phase shifts at these points
\begin{eqnarray} 
&& 
\label{q2}
\hat {M}_c^{(-)} = 
\left
( 
\begin{array}{cc} 
1 & -i \\ 
-(i/2) & (1/2)  
\end{array} 
\right ) 
\, ,
\end{eqnarray}
and $\hat {M}_c^{(+)}$ is the matrix Hermitian conjugated to (\ref{q2}).
The 
$\hat L^{(c)}_{L/R}$ 
and 
$\hat F_c$ 
matrices are called shift matrices, and those are related
to the variations of the coefficients of increasing and decaying semiclassical
solutions in the regions between the turning points ($\hat F_c$ 
is the shift matrix when one moves from the crossing to the turning point in classically
forbidden region, and $\hat L^{(c)}_{L/R}$ 
are the shift matrices in the classically accessible regions).
Explicitely we get
\begin{eqnarray} && 
\label{q3}
\hat F_c =
\left( 
\begin{array}{cc} 
\exp (-\gamma W_B^*/2) & 0 \\ 
0 & \exp (\gamma W_B^*/2) 
\end{array}
\right ) 
\, .
\end{eqnarray}
Here $\gamma $ is the semiclassical
parameter, and $W^*_B$ is the action in the lower adiabatic potential barrier. 
Finally the structure of the shift matrices 
$\hat {L}^{(c)}_{L/R}$ is 
\begin{eqnarray} 
&& 
\label{q33} 
\hat {L}^{(c)}_{L/R} = 
\left( \begin{array}{cc} \exp (i \gamma W_{L/R}^*) & 0 \\ 0 & \exp (-i \gamma
W_{L/R}^*) 
\end{array} 
\right ) \, , 
\end{eqnarray} where $W_{L/R}^*$ is the action calculated by the
integration between the turning points. 
We depicted the 
corresponding trajectories and matrices in Fig. 2

The same manner can be treated the over-barrier region (\ref{j2}) (see Fig. 3).
In this case the crossing point is in the classically accessible region
for the both potentials. The fundamental diabatic solutions can be represented
as the waves propagating in the opposite directions, and the complex-valued
connection matrix has as it was for the tunneling region $2 \times 2$
block structure, where the blocks link the waves in the $L$ and in the $R$
wells propagating in the same direction. Specifically 
the corresponding connection matrix 
at the crossing point $\hat {U}_c^\prime $ 
\begin{eqnarray} && 
\label{q88} \hat {U}_c^\prime =
\left [
\begin{array}{cccc} 
s\exp (-i\phi ) & 0 & 0 & -\exp (-\pi \nu ) \\ 
0 & s \exp (i \phi ) & - \exp (-\pi \nu ) & 0 \\ 
0 & \exp (-\pi \nu ) & s \exp (-i \phi ) & 0 \\ 
\exp (-\pi \nu ) & 0 & 0 & s\exp (i \phi ) 
\end{array} 
\right ] \, , 
\end{eqnarray} 
(where we denoted $ s = \sqrt {1 - \exp (-2\pi \nu )}$,  $\phi = \arg \Gamma (-i \nu )
+ \Im (2 \tilde \chi )$, and $ 
\tilde \chi = - (i/2)((\pi /4) + \nu (1 - \ln \nu)) + 
(1/4)(\pi \nu + \ln \nu )$) 
should
be multiplied by two blocks: the block from the left gives the contribution
at the turning point and includes the shift matrix to the crossing point in $L$
and in $R$ wells of the lower adiabatic potential; the right block is related
to the turning point and to the shift matrix to the crossing point in the
upper one-well adiabatic potential. Thus finally in the over-barrier region we 
get
\begin{eqnarray} && 
\label{q9}
\left ( 
\begin{array}{c} 
\Phi _R^- \\ 
\Phi _R^+ \\ 
\Phi _L^+ \\ 
\Phi _L^- 
\end{array} 
\right ) = 
\left ( 
\begin{array}{cc} 
\hat {M}_c^{(+)}\hat {L}^{(c)}_R & 0 \\ 
0 & \hat {M}^{(+)}\hat L
\end{array}
\right )
\hat {U}_c^\prime  
\left ( 
\begin{array}{cc} 
\hat {L}^{(c)}_L\hat {M}_c^{(-)} & 0 \\ 
0 & \hat {L}\hat {M}^{(-)} 
\end{array} 
\right )
\left ( 
\begin{array}{c} 
\Phi _L^+ \\ 
\Phi _L^- \\ 
\Phi _R^- \\ 
\Phi _R^+ 
\end{array} 
\right ) \, .
\end{eqnarray} 
Here we used the same notations as it was above for the tunneling region, and besides
the matrices 
$\hat M^{(\pm )}$ 
are transposed with respect to the matrices
$\hat M_c^{(\pm )}$ given in (\ref{q2}), and the new shift matrix $\hat L$
is 
\begin{eqnarray} && 
\label{q333}
\left( 
\begin{array}{cc} 
\exp (-i \gamma W^*/2) & 0 \\ 
0 & \exp (i \gamma W^*/2) 
\end{array}
\right ) 
\, ,
\end{eqnarray}
(remind that $W^*$ is the action in the upper adiabatic potential).
Combining altogether (\ref{q9}), (\ref{q88}), (\ref{q333}), and (\ref{q2})
one can trivially find the full connection matrix for the over-barrier energy region (\ref{j2}).

More tricky task is to calculate the connection matrix in the
intermediate energy region (\ref{j3}). In this region the crossing
point is close to the internal linear turning points of the diabatic
potentials. Therefore the two fundamental
diabatic solutions are in the classically accessible region and two others
are in the forbidden region (see Fig. 3 for the illustration). 
Nevertheless even in this case the
connection matrix
has the $2 \times 2$ block structure, but these blocks 
determine the transitions between the adiabatic states
(unlike the tunneling or the
over-barrier regions, where the connection matrices (\ref{q1}), (\ref{q9})
link the diabatic states).
To treat this kind of problems we have developed recently \cite{BV03}, \cite{BV04}
the semiclassical method for level quantization in the vicinity of the
diabatic potential crossing point. The method enables to find all
four exponentially increasing or decreasing solutions to the Schr\"odinger equation
for an arbitrary shape of the crossing diabatic potentials, i.e. for any combination
of the 1-st and 2-d order turning points, and of the crossing point.
In this paper we generalize this method \cite{BV03}, \cite{BV04} to study coherent-incoherent
tunneling regimes in an asymmetric double well potential.

Following the same line as above we first present
the general structure of the connection matrix in the intermediate energy region
\begin{eqnarray} && 
\label{q99}
\left ( 
\begin{array}{c} 
\Phi _R^- \\ 
\Phi _R^+ \\ 
\Phi _L^+ \\ 
\Phi _L^- 
\end{array} 
\right ) = 
\left ( 
\begin{array}{cc} 
\hat {M}_c^{(+)}\hat {L}^{(c)}_R\hat {M}_-^{(+)} & 0 \\ 
0 & \hat {M}_+^{(+)}
\end{array}
\right )
\hat {U}_c^{\prime \prime }  
\left ( 
\begin{array}{cc} 
\hat {M}_-^{(-)}\hat {L}^{(c)}_L\hat {M}_c^{(-)} 
 & 0 \\ 
0 & \hat {M}_+^{(-)}
\end{array} 
\right )
\left ( 
\begin{array}{c} 
\Phi _L^+ \\ 
\Phi _L^- \\ 
\Phi _R^- \\ 
\Phi _R^+ 
\end{array} 
\right ) \, .
\end{eqnarray} 
The matrices $\hat {M}_\pm^{(\pm )}$, have been introduced
in \cite{BV02} for the imaginary turning points (in the case under consideration
the both adiabatic potentials have these turning points in the energy
window $|\alpha | < U_{12}$) and they have a form
$$
\hat {M}_\pm ^{(+)} =
\left [
\begin{array}{cc} 
1 & 0 \\ 
(i/2)\exp (-\gamma W_i^\pm ) & 1 
\end{array} 
\right ] \, , 
$$
where $\hat {M}_\pm^{(-)}$ is the matrix Hermitian conjugated to $\hat {M}_\pm ^{(+)}$,
and $W_i^\pm $ are so-called Euclidean actions in the turned over upper and lower
adiabatic potentials
\begin{eqnarray} 
\label{a55} 
W_i^\pm \simeq \frac{\pi q_{1 , 2}}{\gamma } \, ; \, q_{1 , 2} = \gamma \frac{U_{12}}{4 f}\sqrt {U_{12} \pm \alpha }
, .
\end{eqnarray} 
Since the matrices $\hat {M}_\pm ^{(\pm )}$ become the unitary ones at $\alpha > U_{12}$ and at $\alpha < - U_{12}$,
the intermediate region connection matrix (\ref{q99}) matches continuously the
the connection matrices (\ref{q1}) and (\ref{q9}) in the tunneling and over-barrier regions, respectively.

The
connection matrix in the intermediate energy region can be calculated using the known
Weber function asymptotic expansions
for large complex indices \cite{OL59}, \cite{OL74}.
Combining together these asymptotic expansions 
and all matrices entering (\ref{q99}) defined already above, we find
at the crossing point, the matrix
$\hat U_c^{\prime \prime }$ is
\begin{eqnarray} &&
\label{q888}
\hat {U}_c^{\prime \prime } = \left [
\begin{array}{cc}
(\sqrt {2\pi }/\Gamma (q^*))\exp (-2\chi (q^*)) & 0   \\
0 & (\Gamma (q)/\sqrt {2\pi })\exp (2\chi (q))(1 - \exp (-2\pi q_2)\cos ^2(\pi q_1)) \\
0 & \exp (-2\pi q_2)\cos (\pi q_1)   \\
\exp(-2\pi q_2)\cos (\pi q_1) & 0
\end{array}
\right .
\end{eqnarray}
\begin{eqnarray}
\nonumber
\left .
\begin{array}{cc}
0 & - \exp (-2\pi q_2)\cos (\pi q_1)  \\
-\exp (-2\pi q_2)\cos (\pi q_1)  &
0  \\
(\sqrt {2\pi }/\Gamma (q))\exp (2\chi (q)) & 0 \\
0 & (\Gamma (q^*)/\sqrt {2 \pi })\exp (2\chi (q^*))(1 - \exp (-2\pi
q_2)\cos ^2(\pi q_1))
\end{array}
\right ] \, ,
\end{eqnarray}
where as above we introduced the following abridged notations
\begin{eqnarray} 
\label{a551} 
q  = q_1 + i q_2\, ; \,
q_{1 , 2} = \frac{\gamma u_{12}\sqrt {u_{12} \pm \alpha }}{4 f} \,
; \, q^* = q_1 - i q_2 \, ,
\end{eqnarray}
and, besides, 
\begin{eqnarray}
\label{a552}
\chi = \chi _1 + i \chi _2 \, ; \, 2 \chi _1 = q_1 - \left (q_1 - \frac{1}{2}\right ) \ln |q| + \varphi q_2 \,
,
\end{eqnarray}
and analogously
\begin{eqnarray}
\label{b552}
2\chi _2 = q_2 - q_2\ln |q| - \varphi \left (q_1 - \frac{1}{2}\right )
\, .
\end{eqnarray}
The phase factor $\varphi $ is defined as
\begin{eqnarray}
\label{new}
\tan \varphi = \sqrt {\frac{U_{12} - \alpha }{U_{12} + \alpha }}
\, .
\end{eqnarray}
Now the full connection matrix in the intermediate energy region
can be found easily simply collecting the given above expressions.

\section{Results and discussion}
\label{res}

Our purpose in this section is to study how found in \cite{BK02}
coherent - incoherent tunneling relationships shortly described
in the introduction section \ref{I} for an isolated double-well
potential, and in particularly the criterion (\ref{na1}) and the
dephasing time $T_2$ (\ref{I1}) should be modified
for more realistic situations when there is a finite adiabatic
coupling between the diabatic potentials forming the asymmetric
double-well lower adiabatic potential, and the one-well upper
adiabatic potential.

However to investigate this problem 
first we should derive the quantization
rules for the crossing diabatic potentials. 
It can be done using presented above the connection matrices.
In spite of the fact that instanton trajectories are rather simple objects, and can be relatively easy
found analytically, calculations of the quantization rules within the instanton approach are rather
intricate and require the knowledge of all connection matrices, we
have calculated in the previous section. 
To apply this machinery within the instanton approach, the quantization rule can be formulated 
as a condition that the amplitudes
of exponentially increasing solutions must
be vanished. In terms of the matrix elements $m_{ij}$ of the
connection matrix this condition is
\begin{eqnarray} 
\label{a555} 
m_{22}m_{33} - m_{23}m_{32} = 0
\, . 
\end{eqnarray} 
If made above assumptions are granted one can easily write down
the Bohr - Sommerfeld \cite{LL65} quantization equations applying shown in the Figs. 2, 3
the connection- and shift matrices (for the details of the calculation method, see \cite{BV03},
\cite{BV04}), and we end up with 
\begin{eqnarray} 
\label{a5} 
\tan (\gamma W^*_L)\tan (\gamma W^*_R) = \frac{4}{p^2} \exp (2 \gamma W^*_b)
\, , 
\end{eqnarray} 
where $W^*_b$ is the action in the classically forbidden region in
between the turning points, and $W^*_{L/ R}$ are the coordinate independent
actions inside of the $L$ (respectively $R$) well.
This equation (\ref{a5}) can be solved to find energy levels in the wells.

Applying the same procedure to the over-barrier region (\ref{j2}) we find from
(\ref{a55}) 
\begin{eqnarray} 
\label{a155} 
&&
(1 - \exp (- 2\pi \nu ))\cos (\gamma (W^*_L + W_R^*)- \phi )\cos (\gamma W^* + \phi ) 
+ 
\\
&&
\nonumber
\exp (-2\pi \nu )
\cos \left (\gamma \left (W_L^* + \frac{W^*}{2}\right )\right )
\cos \left (\gamma \left (W_R^* + \frac{W^*}{2}\right )\right ) = 0
\, . 
\end{eqnarray} 
In the diabatic limit ($\nu \to 0$) one get from (\ref{a155})
\begin{eqnarray} 
\label{a255} 
\cos \left (\gamma \left (W_L^* + \frac{W^*}{2}\right )\right )
\cos \left (\gamma \left (W_R^* + \frac{W^*}{2}\right )\right ) = 0
\, , 
\end{eqnarray} 
and therefore two independent quantization conditions
\begin{eqnarray} 
\label{a355} 
\left (\gamma \left (W_L^* + \frac{W^*}{2}\right )\right ) = \pi \left (n_L + \frac{1}{2}\right )
\, ; \,
\left (\gamma \left (W_R^* + \frac{W^*}{2}\right )\right ) = \pi \left (n_R + \frac{1}{2}\right )
\, . 
\end{eqnarray} 
On the other hand in the adiabatic limit, i.e. at $\nu \to \infty $ and $\phi \to 0$
we have
\begin{eqnarray} 
\label{a455} 
\cos (\gamma (W^*_L + W_R^*))\cos (\gamma W^*)=0 
\, , 
\end{eqnarray} 
and therefore
\begin{eqnarray} 
\label{a655} 
\gamma (W^*_L + W_R^*) = 
\pi \left (n + \frac{1}{2}\right )
\, ; \,
(\gamma W^*) = \pi \left (n_0 + \frac{1}{2}\right )
\, . 
\end{eqnarray} 
And to conclude this part and to span a wide
range of possibilities, the quantization condition
in the intermediate energy region derived from the
corresponding connection matrix can be represented in the following
form
\begin{eqnarray} 
\label{n355} 
\cos (\gamma (W^*_L + W_R^*) - \phi ) = \frac{\exp (-\pi q_2)}{\sqrt {1 + \exp (-\pi q_2)}}
\cos (\gamma (W_L^* - W^*_R)) 
\, .
\end{eqnarray} 
Now it seems appropriate to take a fresh look at the results
presented above. What can we learn from the performed calculations?
First we can go one step further to analyze the 
phase factors calculated above.
In our system (two crossing diabatic potentials) there are two types
of phases. The first phase factor 
occurs, since the tunneling results in the
phase shift related to the change of eigenvalues. 
In own turn it leads to a certain kind of one-well phase
($T_2$) relaxation.
The physical argument leading to $T_2$ relaxation at the tunneling
in the asymmetrical double well potentials may be rationalized
as follows.
The fact is that reflected from the barrier waves acquire non trivial phase
factor. The phenomenom is related to interference of incident, 
reflected and transmitted waves. One can look to this phase
factor from a slightly different point of view, since tunneling results in the phase shift related to the
change of eigenvalues. The quantization rules can be rewritten in the 
form which includes some integer numbers
numerating and an exponentially small phase shift due to the existence of the
barrier between two wells. 

The second phase shift occurs in our case due to non-adiabatic behavior. 
Indeed the LZ case (even for the same asymmetric double well shape of the lower adiabatic
potential) is quite different not only quantitatively due to
coupling with the upper adiabatic potential but also qualitatively,
since a novel and fundamental quantum
effect will occur. Namely, in addition
to the described above tunneling phase (existing even in an isolated double well potential)
a quantum mechanical wave function acquires upon a cyclic evolution 
some geometrical or Berry phase
factor \cite{BE} - \cite{FJ98}.
What is most characteristic for the concept of Berry phase is the existence of a continuous parameter
space in which the state of the system can travel on a closed path. In our case the phase is determined by
the non-adiabatic interaction. 
Coherent or incoherent kind of behavior for  
crossing diabatic potentials crucially depends on a quite tricky interplay
between the both (i.e. tunneling and Berry) phase factors.
Two new results which have emanated from our study of these phenomena, is our main motivation for
presenting
this paper.
The semiclassical wave functions of the bound states are linear combinations
of (\ref{q1}), or (\ref{q9}), or (\ref{q99}) (respectively for the tunneling, over-barrier,
and intermediate energy regions), which can be determined, provided we know the eigenvalues.
The quantization rules (namely, conditions that coefficients at exponentially
increasing and ingoing from the both infinities $|X| \to \pm \infty $ 
waves are zero),
and the wave function normalizations, define uniquely these linear combinations. 
The shift matrices have always the same form as (\ref{q33}), where one has to insert the action between
a given point $X$ and the nearest turning (or crossing) point $X_k$. For the upper and the lower
adiabatic potentials the action reads as
\begin{eqnarray} 
\label{new1} 
W_\pm (X_k , X) = \int _{X_k}^{X} d X \sqrt {2 \left |U^\pm (X) - \frac{E}{\gamma }\right |} 
\, .
\end{eqnarray} 
We have illustrated the general connection matrix scheme in the Figs. 2 and 3.

Let us consider a general example describing
two non-symmetric potentials crossing at $X = 0$ point (\ref{nn1}).
When the parameter $b$ entering
the potential (\ref{nn1}) is varied from 1 to $\infty $, we recover 
the two known in the literature limiting cases,
and come from two identical parabolic potentials 
to the case one-well and linear diabatic 
potentials crossing. 
This kind of the diabatic potentials
crossing leads to the lower adiabatic potential in the
form investigated in \cite{BK02}, and has qualitatively
the same features as the model potential (\ref{a2}).
If one neglects for a moment the upper adiabatic potential,
aiming to study crossover behavior from
coherent to incoherent
tunneling upon increase of the parameter $b$, the larger is this parameter
$b$, the larger will be the density of final states.
The criterion for coherent-incoherent crossover behavior found in 
\cite{BK02} based on comparison of the transition matrix elements and
the inter level spacings in the final state. The analogous
criterion should hold for LZ level crossing problem, however in the latter
case the tunneling transition matrix elements has to be multiplied by the 
adiabatic factor. Therefore the coherent - incoherent tunneling
crossover region moves towards the more dense density of final states,
and the larger $U_{12}$ is the smaller will be the region for
incoherent tunneling.
For the sake of the skeptical reader it is worth to emphasize that the 
tunneling matrix element dependence on the Massey parameter $\nu $
found above is valid for an arbitrary magnitude of the adiabatic
coupling (cf. the recent publication \cite{SK02} where this matrix element
has been calculated in the frame work of the perturbation theory,
and only in the adiabatic and diabatic limits).

Owing to the non-adiabatic behavior of the system the tunneling matrix
element $H_{LR}$ is renormalized by the adiabatic factor. In the
tunneling region from (\ref{a5}) we find 
this renormalization as
\begin{eqnarray}
\label{z1}
H_{LR} \to H_{LR} p(\nu )
\, ,
\end{eqnarray}
where the function $p(\nu )$ (\ref{xx})
is associated with the transition amplitudes between the diabatic
potentials in the crossing region.

This renormalization tunneling factor varies from 0 to 1 upon increasing 
of the Massey parameter $\nu $. Again as we have found
for the isolated double-well potential, in the limit
\begin{eqnarray}
\label{z11}
H_{LR} p(\nu ) \ll \Delta _R
\, ,
\end{eqnarray}
the spectrum consists of the set of the tunneling doublets
and $L - R$ transitions are coherent ones. This criterion (\ref{z11})
replaces (\ref{na1}) for our case of the finite adiabatic coupling
$U_{12}$.

Quite different situation occurs for the excited states. In the diabatic limit, 
the transition
matrix element is increased with the Massey parameter $\nu $, and therefore
at a given $b$ value, the system moves to more incoherent behavior.
In the adiabatic limit, the transition matrix element is exponentially small,
and coherence of the inter-well transitions should be restored.
However, since the matrix elements are oscillating functions of $U_{12}$
for the intermediate range of this coupling ($U_{12}$) coherent - incoherent
tunneling rates are also non-monotonically varying functions.
To illustrate it let us study dynamics of the initial state of the system,
prepared somehow in the ground vibrational state of the upper adiabatic
potential
\begin{eqnarray}
\label{ben1}
U^{(+)} = \frac{1}{2} + \frac{1+b}{4b} X^2 + \left [U_{12}^2 + X^2\left (1 + \frac{b-1}{4}X\right )^2\right ]^{1/2}
\, .
\end{eqnarray}
Evidently the wave function $\psi ^{(0)}$ of this state
should be close to the harmonic oscillator function $n=0$, with its minimum at 
$X=0$, and its eigen-state $E_0$ dependent of $U_{12}$. The real part of the $E_0$
gives the oscillator vibration
frequency $\omega _0$, and the imaginary part determines the decay rate $\Gamma _0$ of the
state.

The spectral expansion of $\psi ^{(0)}$ over the diabatic
state eigen-functions $\{ \psi _n\} $ can be found by the method proposed
long ago by Zeldovich \cite{ZE61} (see also \cite{BK02}, where the method
has been applied to
find semiclassical solutions of the Schr\"odinger equation for
1D asymmetric double well potential)
well adapted for quasi-stationary state wave function expansion over
continuum spectrum functions. Note that for $b \gg 1$ all the states of
the upper adiabatic potential (\ref{ben1}) are placed in the same energy range
that the $R$-well excited states. Since, as it is well known \cite{LL65}, the harmonic oscillator
wave functions for $n \gg 1$ coincide with the semiclassical ones, we can represent
the $R$-well wave functions in the vicinity of $E_0$ in the following form 
\begin{eqnarray}
\label{ben2}
\psi _n(X) =
\left \{
\begin{array}{c}
A(k_n) \psi _n^{(0)}(X) \, , \, |X| \simeq \lambda _0 \\
\sqrt {\frac{2}{\pi }}\sin (k_nX + \varphi (k_n)) \, , \, X \gg \lambda _0
\end{array}
\right .
\, ,
\end{eqnarray}
where $\lambda _0 \equiv \hbar /m\omega _0$ is the de Broglie wave length,
corresponding the ground state oscillator wave function, $\psi _n^{(0)}$ are harmonic oscillator
eigen - functions, $k_n = \sqrt {2mE_n/\hbar }$ is the wave vector, and the phase
factor $\varphi (k_n)$ is defined as
\begin{eqnarray}
\label{ben3}
\varphi (k_n) = \frac{k_0^{\prime \prime }}{k_n - k_0^\prime }
\, , \, k_0^\prime = \frac{\sqrt {2mE_0}}{\hbar } \, , \, k_0^{\prime \prime } = k_0^\prime \frac{\Gamma _0}{4 E_0}
\, .
\end{eqnarray}
The entering (\ref{ben2}) amplitudes $A(k_n)$ are determined by the condition
that the probability density flow from the quasi-stationary state to infinity
should be constant (in fact this condition
plays the role of the normalization condition for the quasi-stationary states \cite{ZE61}):
\begin{eqnarray}
\label{ben4}
A^2(k_n) = \frac{2 \hbar }{\pi }\sqrt {\frac{2 E_n}{m}} \frac{\Gamma _0}{4 (E_n - E_0)^2 + \Gamma _0^2}
\, .
\end{eqnarray}
We have shown in the Fig. 4 (for a fixed value of the potential shape controlling parameter
$b \gg 1$ and various coupling strengths $U_{12}$) the spectral density
expansion
\begin{eqnarray}
\label{ben5}
S(E) = \sum _{n}|<\psi |\psi _n>|^2\delta (E - E_n)
\, .
\end{eqnarray}
The spectrum of final states for $ b \gg 1$ is a discrete one (although dense), and
the envelope of the spectrum has
a Lorentzian shape with a width $\Gamma _0$, which is determined
by the non-adiabatic transition matrix element. The latter quantity has
an oscillating dependence on $U_{12}$. Since the final state spectrum at $b \gg 1$
has only weak dependence on the non-adiabatic coupling $U_{12}$ we have almost
constant spectral distribution but the number
of the final states (relevant for the transition) is determined by $\Gamma _0$, and
it oscillates with $U_{12}$. Analogously to the adiabatic transitions considered
in \cite{BK02}, one can formulate the non-reversibility criterion for non-adiabatic
transitions. Indeed coherent oscillations turn into exponential decay when sufficiently
large number of the final states occurring under the Lorentzian envelope, i.e., when
\begin{eqnarray}
\label{ben51}
\Gamma _0 \gg \Delta E_n
\, .
\end{eqnarray}
It is worth noting that unlike the adiabatic transitions \cite{BK02},
when $\Gamma _0$ is determined uniquely by the amplitudes of the inter - well
transitions, in our case for the quasi-stationary states of the upper adiabatic
potential, $\Gamma _0$ depends on the rate of the non-adiabatic transitions.
To calculate the rate one has to solve first the eigenvalue problem.

The spectral distribution (\ref{ben5}) fully characterizes the state $\psi (t)$
time evolution. In particularly relevant quantity is so-called survival probability
$P(t)$ reads as
\begin{eqnarray}
\label{ben52}
P(t) = \left |\int _{-\infty }^{+\infty } d E S(E) \exp (i E t/\hbar ) \right |^2 
\, ,
\end{eqnarray}
and it is presented in the Fig. 5.
We see that there are a number of the coupling strengths $U_{12}$ regions with
the non-monotonous behavior. The coherent dynamics holds for the energy regions
not too close to the $L$ well levels, when only few $R$ states coupled to the state
$\psi (t)$. 
In the case the energy of the state is in the resonance with
a certain level of the $L$ well, the transition matrix elements are enhanced. In this case
the $\psi (t)$ evolution resembles decays of the quasi-stationary states (see e.g., \cite{BV03},
\cite{BV04}). It is convenient and instructive to illustrate the non-monotonous behavior
by averaging of the surviving probability over the period of the initial state recovering.
Evidently this average value $<P>_T$ equals 1/2 for the coherent oscillations.
The dependence of $<P>_T$ on $U_{12}$ is presented in the Fig. 6.
Note that in the both (adiabatic and diabatic) limits we have coherent oscillations,
as it should be (although over very different time scales), while
in the intermediate energy region the coherent evolution is multiply destroyed
in the vicinity of the resonances.

\section{Conclusion.}
\label{con}

To conclude let us comment first on our motivation.
In principle potentials with two stable equilibrium configurations
are widely used in chemistry and physics, to describe molecular
spectroscopy data. 
Analyzing these data one must distinguish two types
of states which require a set of different theoretical and experimental
methods, each one with specific strengths and weaknesses on certain scales
of length and time.
Experimentally low-energy states localized near the minima of such potentials,
are studied by vibrational spectroscopy methods. These low-energy
states can be characterized by well defined quantum numbers describing
the normal vibration excitations. 
Quite different
approaches are necessary to use to study highly excited states situated
near the potential barrier top. Just these states determine the probability
of thermo-activated molecular transitions. These phenomena are intrinsically
statistical ones, and not surprisingly theoretical descriptions of these
states assume usually their ergodicity. 
This ergodic behavior can
be easily understood since the excited states near the barrier top
have so high density that even very small coupling to an environment
(thermal reservoirs) can provide say fast mixing and thermalization of the
states.

However applying these approaches (and the model potentials) to real 
chemical dynamic problems of low-temperature reactions and transitions
of relatively small molecules or atomic clusters (attracting much attention
in relations with chemical reactions in upper Earth atmosphere layers,
and high precision laser spectroscopy techniques),
one should take care whether these two relevant regions of energy
are not overlapped.
Measurements of molecules with two stable configurations
performed in the temperature interval ($10 - 20 \, K$) low enough
to provide that for the measurement time dephasing
or relaxation processes are not essential \cite{v1}, \cite{v2}, \cite{BM94},
demonstrated tunneling doublets dependence on well defined vibrational excitations.
Thus this low-temperature behavior can be attributed with the coherent 
tunneling, and the advent of ultrafast lasers has provided
physical chemists with a tool for studying these systems under nonequlibrium
conditions.

On the other hand 
there are also numerous examples (see e.g. \cite{v4}, \cite{v5}, \cite{v6}, \cite{v7})
clearly showing exponential (incoherent) decay of an initially
prepared seemingly equilibrium configuration.
Thus experimental data signal that two distinct
dynamic regimes exist in bistable molecular systems. Moreover, barrier
heights and potential well asymmetries in the systems manifesting different
dynamic behaviors, are quite similar by their magnitudes (of the same order).
A question of primary importance is the understanding of how these
two tunneling regimes (coherent or incoherent) depend on more subtle
specific features
of the potential energy profiles than merely energies of the characteristic points 
(like for example, our model potential (\ref{a2}), (\ref{nn1})).
One qualitative answer to this question has been done long ago by Jortner and Bixon
\cite{v8}. In this paper the authors have formulated
the irreversibility criterion. According to the criterion, coherent tunneling
should be destroyed when the density of final states is so high that 
typical inter-level spacings become smaller than characteristic transition
matrix elements.
Our aim in this paper is to formulate quantitatively the analogous criterion.
One methodical comment seems in order here. Usual technique to analyze
radiationless transitions is based on the perturbation theory in the adiabatic
representation, and the non-adiabatic coupling operator is treated as a perturbation
(see, e.g., \cite{NU84}, \cite{BL00}).
It is easy to understand, however, that this kind of the adiabatic perturbation theory
is equivalent to isolated two level systems approach, which is valid only when
the level displacements (due to tunneling or non-adiabatic transitions) are smaller than the
inter-level spacing. Analogously for quasi-stationary states
the adiabatic perturbation theory works when the level broadening (or decay rate)
is smaller that the level spacing. Clearly it is not the case for the intermediate energy
region we have studied in our paper. Note also that our connection matrix approach can be also formulated
in terms of the Liouville - von Neumann equation for the density matrix, where so-called relaxational
matrix should be chosen phenomenologically to mimic decay rate dependences on the energy and
on the Massey parameter. But anyway to find those one has to solve the Schr\"odinger equation
for the potential under consideration.

Our model appears to be a simplest one demonstrating that relatively small
variation of the adiabatic coupling (at a level spacing scale which is
small in comparison with potential barrier heights or well asymmetries)
enables to change qualitatively dynamic behavior. Therefore, we conclude
that dynamic irreversibility in the systems under investigations
crucially depends on the final states density (and not on
potential energy profiles directly).
We believe we are the first to explicitely address this issue.
To illustrate these phenomena we investigated  
coherent and incoherent tunneling 
in the conditions of crossing diabatic potentials. 
As a result of the coupling and level crossing avoiding, we get the asymmetric
double-well lower adiabatic potential with a variable shape 
depending on a value of a continuous parameter $b$ (which describes in the limit
$b=1$ two identical parabolic diabatic potential crossing and in the limit
$b \to \infty $ one-well and linear diabatic
potentials crossing).                                                                                         
The doublet structure
of levels (generic for double-well potentials) is remained valid as
long as the renormalized by the adiabatic coupling
transition matrix element $H_{LR}$ (or tunneling
splitting)
is smaller than characteristic inter-level spacings $\Delta _R$
We calculated the non-adiabatic factor, and found in the diabatic limit 
($U_{12} \to 0$) 
$H_{LR}$ goes to zero, and
in the adiabatic limit ($U_{12} \to \infty $) 
the tunneling
transitions do not depend on the upper potential.
In the over-barrier energy region $H_{LR}$ is an oscillating function
of $U_{12}$, due to the resonances between the states in the lower and in the upper
adiabatic potentials.
In the case $H_{LR} > \Delta _R$, any level from the shallow $L$-well is coupled
by the tunneling to several levels in the $R$-well,
and the transitions lose their coherence.

In an apparently unrelated development researches studying the
problem of intermolecular energy redistribution discovered purely
quantum energy flow between modes which would be otherwise uncoupled.
The mechanism for such classically forbidden energy flow between degenerate vibrational
modes arose from non-adiabatic couplings involving a sequence
of intermediate states. In the various existing in the nature molecular systems
the non-adiabatic coupling strengths $U_{12}$ can have fairly different magnitudes,
thus we anticipate realizations of the both kinds of dynamic behavior, coherent and incoherent
ones. Our model of the non-adiabatic transitions from the initially prepared quasi-stationary
state $\psi (t)$ can be directly confronted to experimental data on super fast non-linear
optical spectroscopy (see e.g., the monographes \cite{FL86}, \cite{MU95}).
The developed in this area technique allows to prepare a given initial
quasi-stationary state by a suitable optical pumping pulse shape.

\acknowledgements 
The research described in this publication was made possible in part by RFFR Grants. 
One of us (E.K.) is thankful to INTAS Grant (under No. 01-0105) for partial support,
and V.B. and E.V. are indebted to CRDF Grant RU-C1-2575-MO-04.

\newpage

\centerline{Figure Captions.}

Fig. 1

Crossing asymmetric parabolic diabatic potentials (adiabatic potentials for
$U_{12} = 0.5$) are shown by the dashed lines):
 
(a) bound initial and final states ($b=3$);
 
(b) bound initial and decay final states ($b = \infty $).

Fig. 2

Connection matrices for the tunneling energy region:
 
(a) in the WKB approach, where the trajectory has 4 linear turning points, and one crossing point.
$M_c^{(\pm )}$ are the connection matrices for the isolated linear
turning points, 
$L^{(c)}_{L/R}$ are the shift matrices (\ref{q33}) in the
classically accessible regions, $F_c$ 
is the shift matrix (\ref{q3}, and $U_c$ is the connection
matrix at the crossing point;

(b) in the instanton method, where the trajectory $E=0$ passes through the second order
turning point (the $L$ well potential minimum). In the $L$ well the WKB connection matrix
should be replaced by the energy dependent connection matrix $M_L^{(2)}$ at the second order
turning point \cite{BV02}), corresponding to kink-anti-kink pair. The same manner the connection matrix for the $R$ well
$M_R^{(2)}$ corresponds to so-called bounce \cite{BV99}.

Fig. 3

Connection matrices for the over-barrier and intermediate energy regions (in the latter region
the connection matrices for the imaginary 
turning points are not shown).

Fig. 4 

Spectral distributions for the initial state wave function (chosen as the ground state of the upper
adiabatic potential) over the eigenfunctions of the model potential (\ref{a2}):
$b=1500$, $\gamma = 12$, and  $U_{12} = 0.09\, , \, 0.15 , \, 0.21 \, , \, 0.28 \, , \,
0.40 $ for (a) - (e) figures respectively.

Fig. 5
 
Survival probability for the same as in Fig. 4 initial state. 

(a) - solid line corresponds to the Fig. 4a; dashed line - to the Fig. 4b;

(b) solid line corresponds to the Fig. 4c, dashed line - to the Fig. 4d.

Fig. 6

Averaged over the recovering period survival probability shown in Fig. 5.

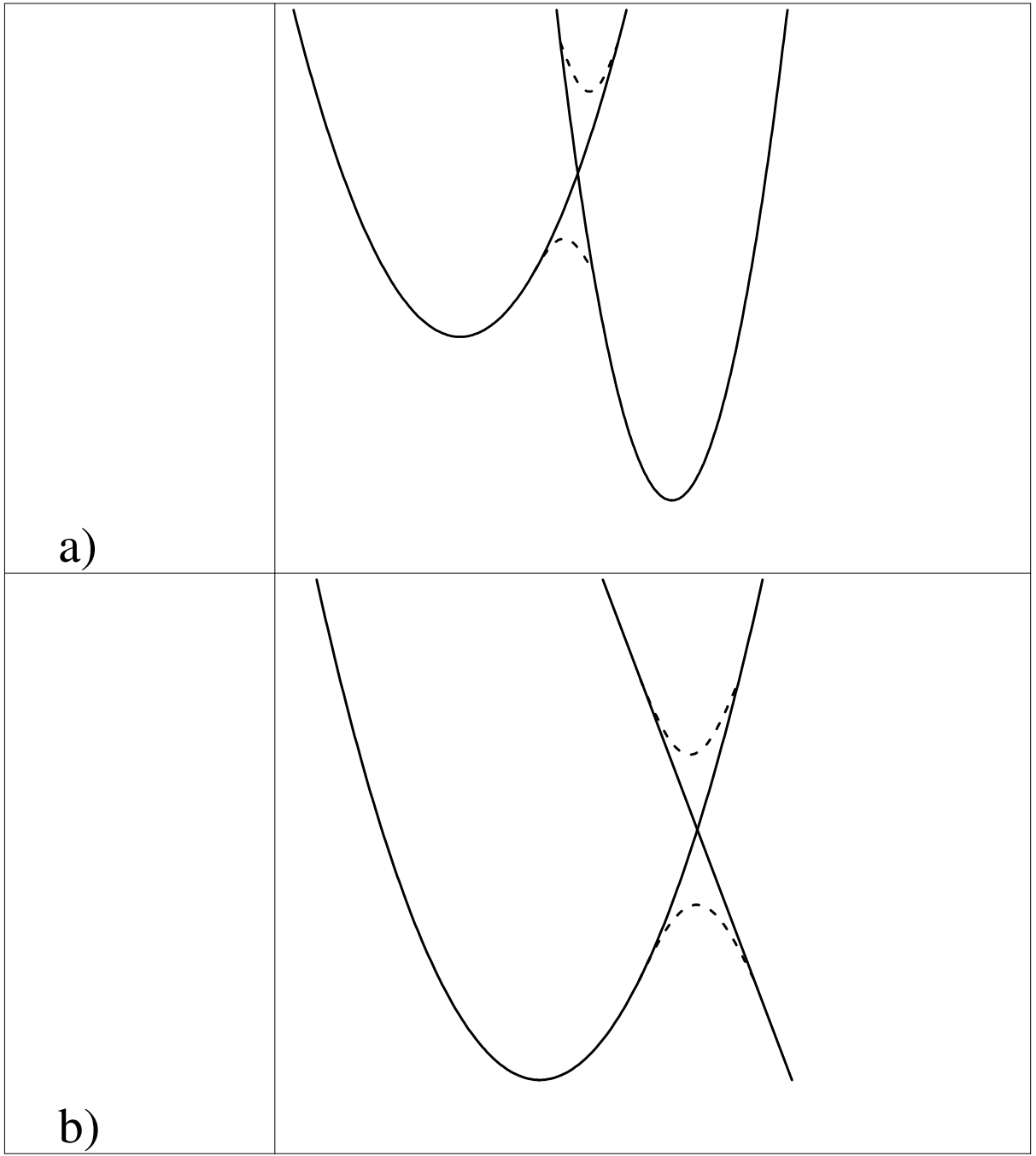
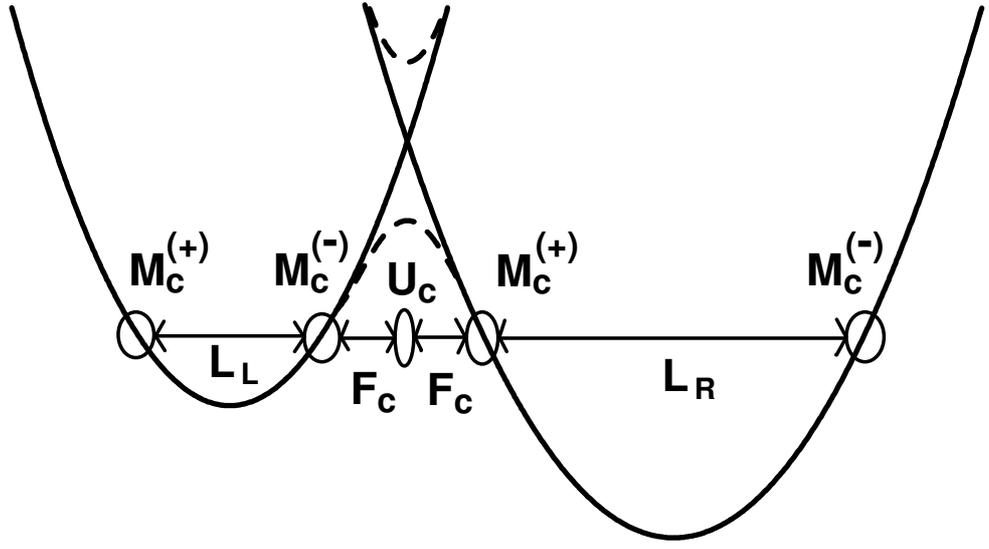
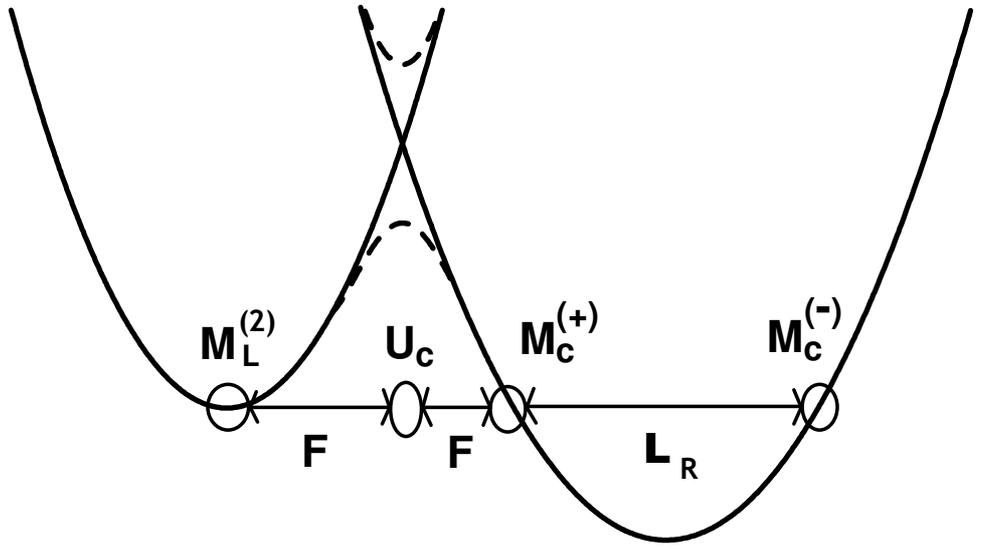
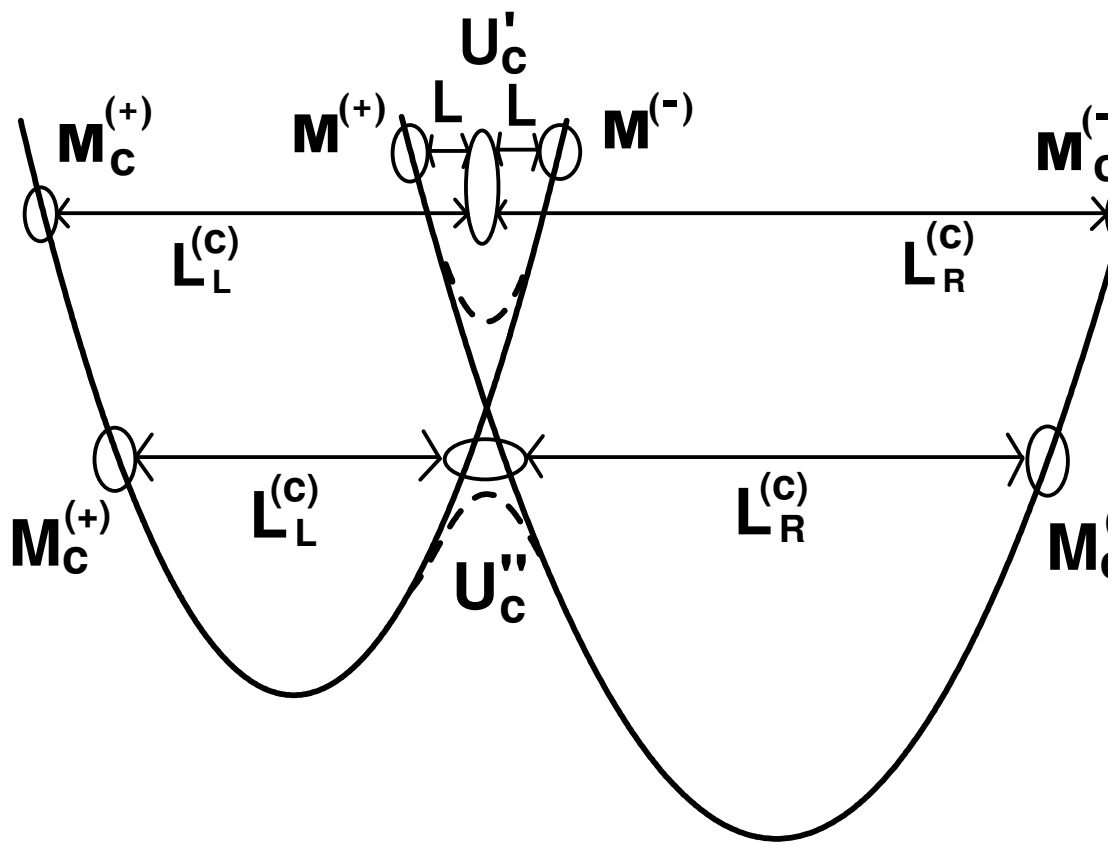
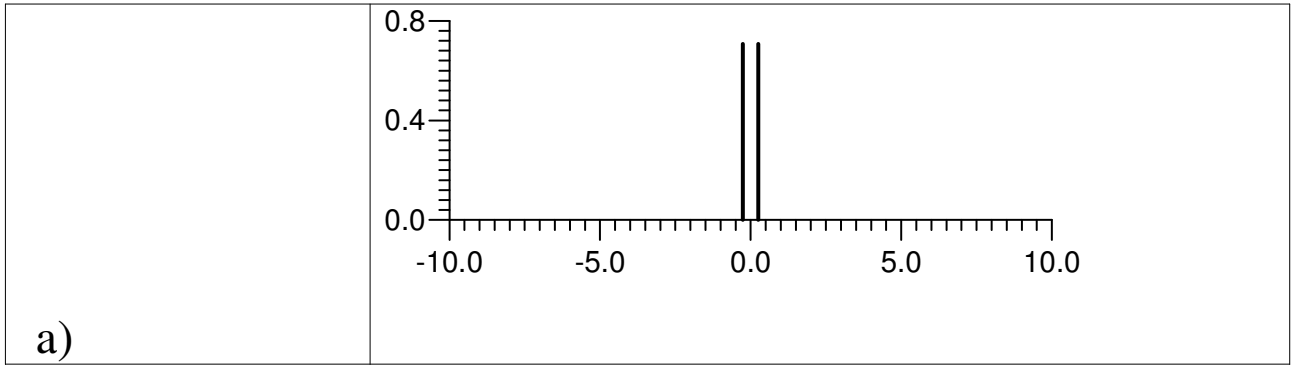
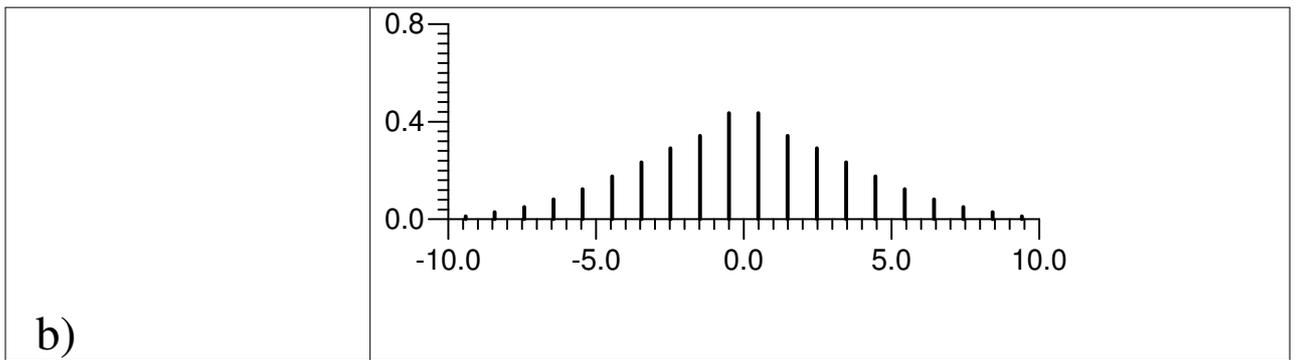
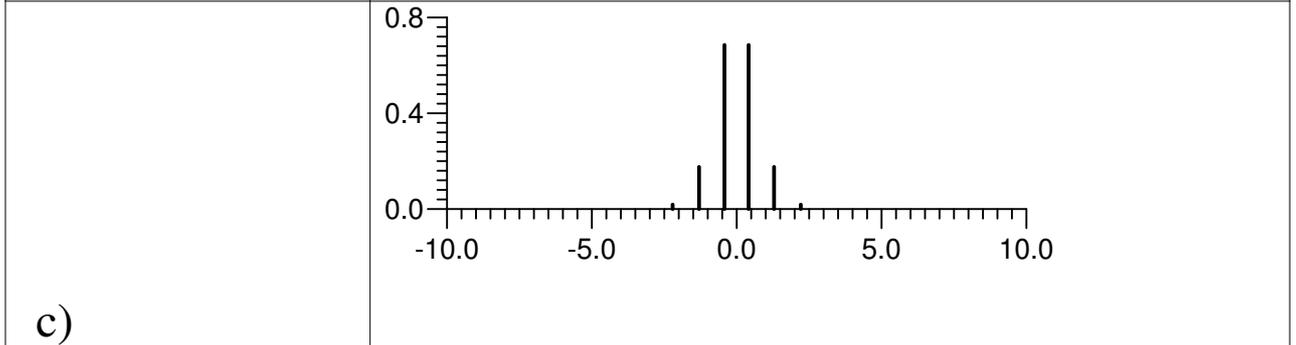
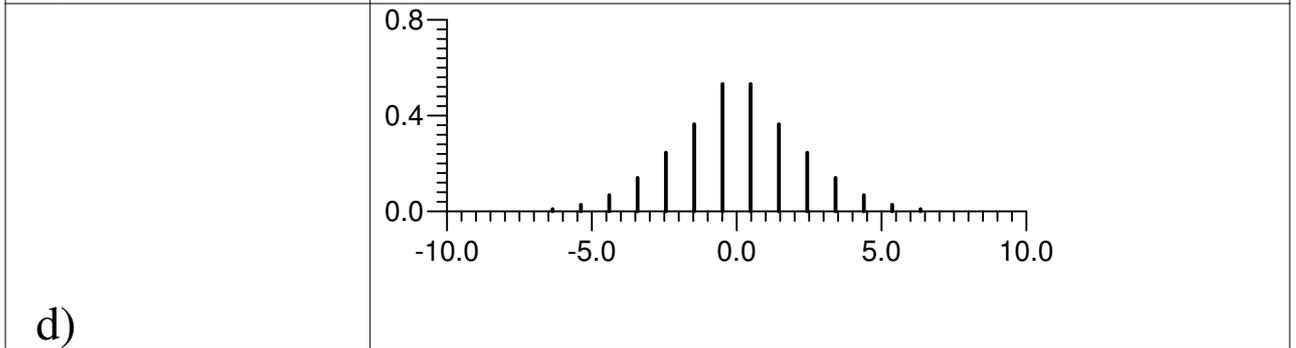
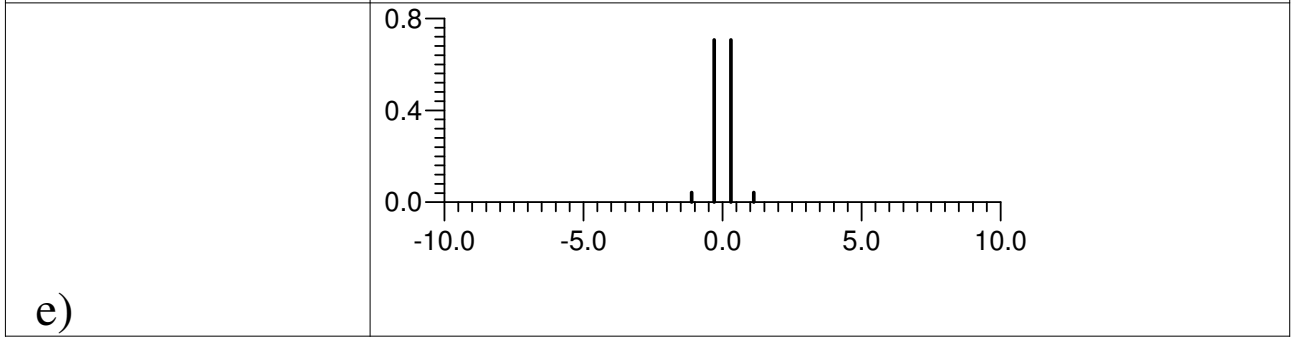
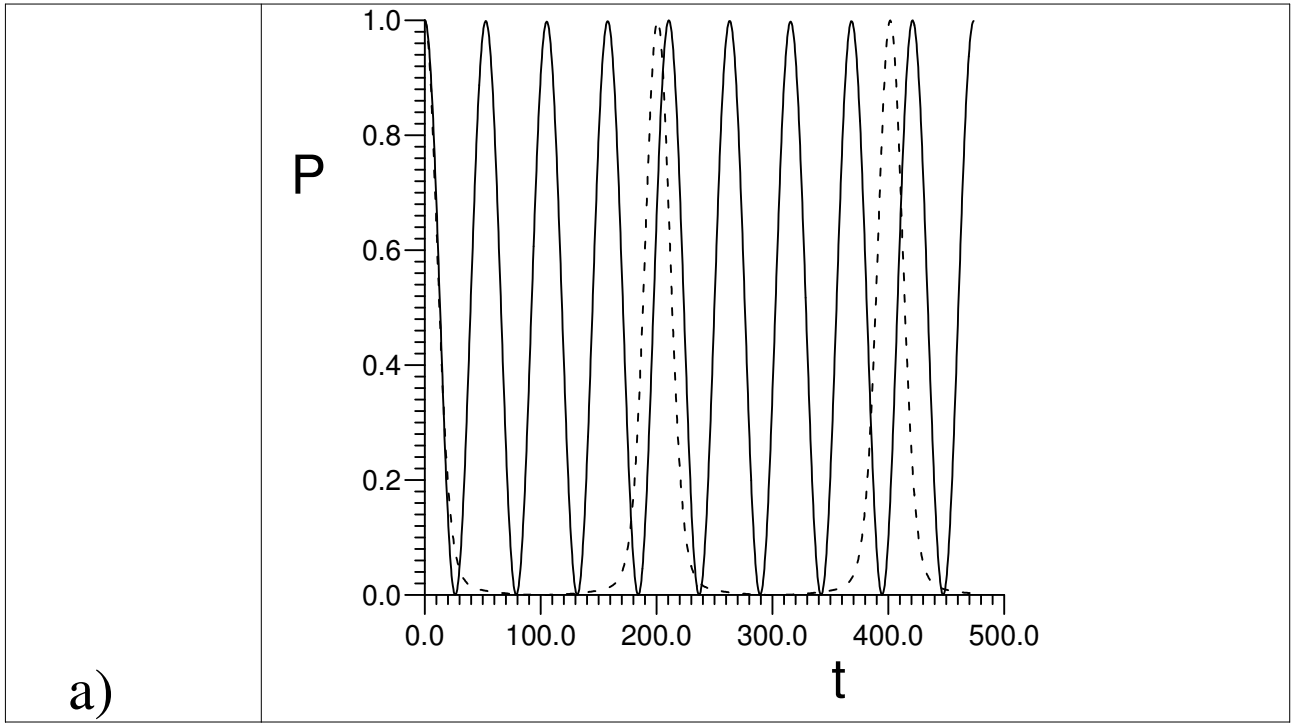
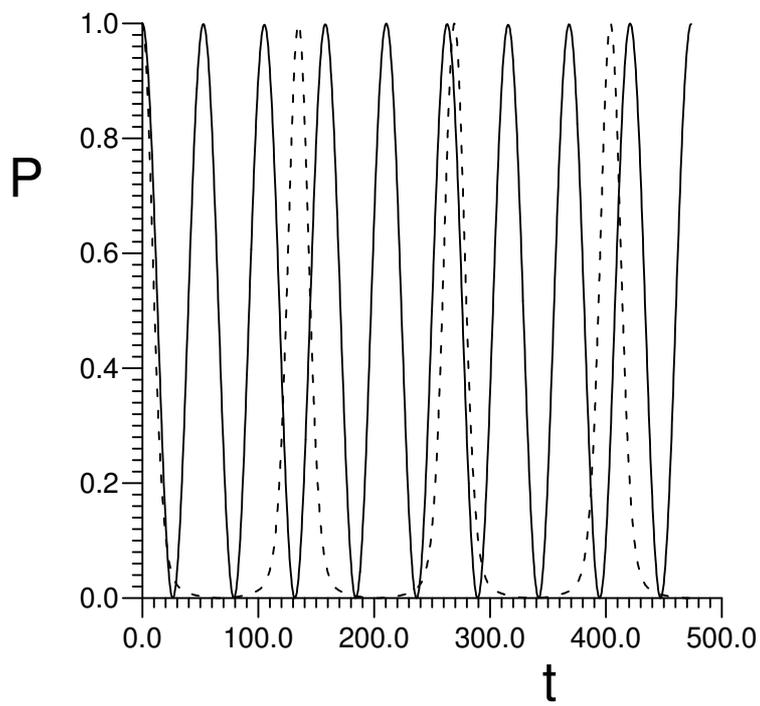
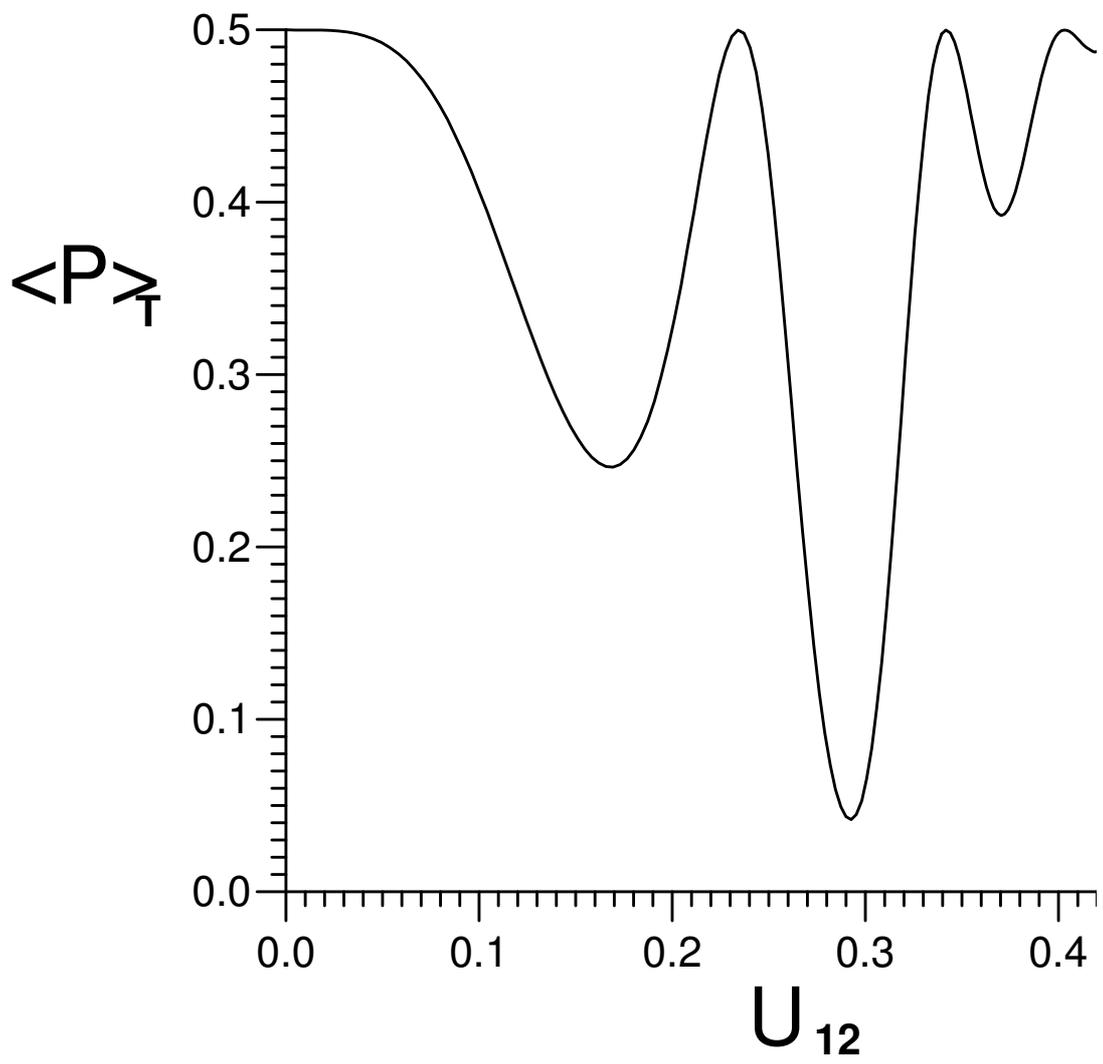
\end{document}